\documentclass[runningheads,a4paper]{llncs}

\usepackage{graphicx}
\usepackage{color}

\usepackage{amsthm}
\usepackage{graphicx,rotating}
\usepackage{complexity}
\usepackage{tikz}
\usepackage{xspace}
\usepackage{fix-cm}
\usetikzlibrary{shadows}
\usetikzlibrary{shapes,positioning,arrows}
\usetikzlibrary{decorations.pathmorphing}
\usetikzlibrary{shadows,arrows,patterns,automata,calc,intersections,fit,shapes.misc, decorations.markings}
\usepackage[underline=false,rounded corners=true]{pgf-umlsd}
\usepackage{todonotes}
\usepackage{paralist}
\setdefaultenum{1)}{(a)}{i)}{A)}

\usepackage[
bookmarks=false,
breaklinks=true,
colorlinks=true,
linkcolor=black,
citecolor=black,
urlcolor=black,
pdfpagelayout=SinglePage
]{hyperref}
\usepackage[all]{hypcap}

\usepackage{dsfont}
\usepackage[bottom]{footmisc}
\usepackage[nocompress]{cite}
\usepackage{listings}
\usepackage{fancyvrb}
\usepackage{cleveref}
\usepackage{bm}
\usepackage{amssymb}
\usepackage{amsmath}
\usepackage{mathtools}
\usepackage{etoolbox}
\usepackage{varwidth} 
\usepackage{array} 
\usepackage{chngcntr}
\usepackage{stmaryrd}

\usepackage{wrapfig,epsfig}

\usepackage{xcolor}
\usepackage{microtype}
\usepackage[T1]{fontenc}
\usepackage{textcomp}

\usepackage{enumitem}

\usepackage{eventB}
\usepackage{zed-csp}

\usepackage[linesnumbered, ruled, noend]{algorithm2e} 
\usepackage{algorithmicx}

\usepackage{subfig}
\usepackage{float}
\usepackage{listing}
\usepackage{mdframed}

\tikzstyle{bdd-leaf}=[circle,draw,solid,thick,minimum size=6mm, inner sep=0mm]
\tikzstyle{ldd-var-two}=[rectangle,draw, solid,minimum size=6mm, inner sep=0mm, rectangle split, rectangle split parts=2, rectangle split horizontal, align=center, text width=.6cm]
\tikzstyle{ldd-var}=[rectangle,draw, solid,minimum size=6mm, inner sep=0mm]
\tikzstyle{ldd}=[node distance=2cm, every edge/.style={-stealth, draw}]
\tikzstyle{normal}=[out=270, in=90]
\tikzstyle{false}=[out=0, in=90, max distance=1.9cm]


\lstdefinestyle{multiline}{
basicstyle=\scriptsize\sffamily,
numbers=left,
numberstyle=\tiny,
frame=tb,
columns=fullflexible,
showstringspaces=false,
escapeinside=\`\`
}

\let\bbordermatrix\bordermatrix
\patchcmd{\bbordermatrix}{8.75}{4.75}{}{}
\patchcmd{\bbordermatrix}{\left(}{\left[}{}{}  
\patchcmd{\bbordermatrix}{\right)}{\right]}{}{}

\let\abordermatrix\bordermatrix
\patchcmd{\abordermatrix}{8.75}{4.75}{}{}
\patchcmd{\abordermatrix}{\left(}{\left\langle}{}{}
\patchcmd{\abordermatrix}{\right)}{\right\rangle}{}{}
\usepackage{etoolbox}

\AtBeginDocument{%
  \mathchardef\mathcomma\mathcode`\,
  \mathcode`\,="8000 
}
{\catcode`,=\active
  \gdef,{\mathcomma\discretionary{}{}{}}
}

\newcolumntype{M}{>{\begin{varwidth}{1.5cm}}l<{\end{varwidth}}} 
\makeatletter
\newcommand{\raisemath}[1]{\mathpalette{\raisem@th{#1}}}
\newcommand{\raisem@th}[3]{\raisebox{#1}{$#2#3$}}
\makeatother

\counterwithin*{equation}{definition}
\theoremstyle{remark}

\theoremstyle{remark}

\numberwithin{mysubcase}{mycase}

\newcommand{\Fref}[1]{Figure~\ref{#1}}

\def\setminus{-}

\def\Btrue{\;\mbox{\bf true}}
\def\Bfalse{\;\mbox{\bf false}}
\def\Bany{\;\mbox{\bf any}\;}
\def\Bwhere{\;\mbox{\bf where}\;}
\def\Bwhen{\;\mbox{\bf when}\;}
\def\Bthen{\;\mbox{\bf then}\;}
\def\Bend{\;\mbox{\bf end}}
\def\Brefines{\;\mbox{\bf refines}\;}
\def\Bstatus{\;\mbox{\bf status}}

\restylefloat{figure}

\crefname{algocf}{alg.}{algs.}
\Crefname{algocf}{Algorithm}{Algorithms}
\crefalias{AlgoLine}{line}
\makeatletter
\let\cref@old@stepcounter\stepcounter
\def\stepcounter#1{%
  \cref@old@stepcounter{#1}%
  \cref@constructprefix{#1}{\cref@result}%
  \@ifundefined{cref@#1@alias}%
    {\def\@tempa{#1}}%
    {\def\@tempa{\csname cref@#1@alias\endcsname}}%
  \protected@edef\cref@currentlabel{%
    [\@tempa][\arabic{#1}][\cref@result]%
    \csname p@#1\endcsname\csname the#1\endcsname}}
\makeatother

\def\Bany{\;\mbox{\bf ANY }}
\def\Bwhere{\;\mbox{\bf WHERE }}
\def\Bbegin{\;\mbox{\bf BEGIN }}

\def\Bthen{\;\mbox{\bf THEN}\;}
\def\Bend{\;\mbox{\bf END}}

\providecommand{\tuple}[1]{\left( #1 \right)}
\providecommand{\set}[1]{\left\lbrace #1 \right\rbrace}
\providecommand{\set}[1]{\left\lbrace #1 \right\rbrace}

\newcommand{\suchthat}{\mathrel{.}}
\newcommand{\impl}{\ensuremath{\mathop{\Longrightarrow}}}

\newcommand{\pins}{P\textsc{ins}\xspace}
\newcommand{\nextstate}{\textsc{NextState}\xspace}
\newcommand{\pinsfull}{\textsc{\underline{P}artitioned \underline{I}nterface} to the \underline{N}ext-\underline{S}tate function \xspace (\pins)}

\newcommand{\ltsmin}{\textsc{LTSmin}\xspace}   
\newcommand{\prob}{\textsc{ProB}\xspace} 
\newcommand{\tlc}{\textsc{Tlc}\xspace}
\newcommand{\tla}{\textsc{Tla}$^{+}$\xspace}

\newcommand{\pinspins}{\textsc{Pins}2\textsc{Pins}\xspace}

\newcommand{\natnumset}{\mathbb{N}}

\newcommand{\mcrl}{m\textsc{crl}2\xspace}

\newcommand{\promela}{\textsc{Promela}\xspace}   

\let\emptyset\varnothing

\newcommand{\numrows}{\ensuremath{\textsc{m}}\xspace}
\newcommand{\numcols}{\ensuremath{\textsc{n}}\xspace}

\newcommand{\pts}{\ensuremath{\mathcal{P}}\xspace}

\newcommand{\initse}{\ensuremath{I}}
\newcommand{\inits}{\ensuremath{I^\numcols}}

\newcommand{\transrel}{\ensuremath{\mathop{\to}}}
\newcommand{\transclose}{\ensuremath{\mathop{\to^*}}}
\newcommand{\transrels}{\ensuremath{\mathop{\to^\numrows}}}
\newcommand{\statese}{\ensuremath{\mathord{S}}}
\newcommand{\states}{\ensuremath{\mathop{S^\numcols}}}

\newcommand{\transrelexpr}{\mathord{\hookrightarrow}}

\newenvironment{zitemize}
     {\begin{list}{--}{
     \setlength{\itemsep}{0 pt}
     \setlength{\parsep}{0 pt}
     \setlength{\topsep} {0 pt} }}
     {\end{list}}
\newcommand{\ignore}[1]{}

\newcommand{\condtext}[2]{#2}

\algnewcommand{\algorithmicgoto}{\textbf{go to}}%
\algnewcommand{\Goto}[1]{\algorithmicgoto~\ref{#1}}%

\makeatletter
\renewcommand*{\@fnsymbol}[1]{\ensuremath{\ifcase#1\or *\or \dagger\or \ddagger\or
   \mathsection\or \mathparagraph\or \|\or **\or \dagger\dagger
   \or \ddagger\ddagger \else\@ctrerr\fi}}
\g@addto@macro{\UrlBreaks}{\UrlOrds}
\makeatother

\begin{document}

\title{Symbolic Reachability Analysis of B through \prob and \ltsmin}
\titlerunning{Symbolic Reachability Analysis of the B-Method}

\author{
Jens Bendisposto\inst{1} \and%
Philipp K\"orner\inst{1} \and%
Michael Leuschel\inst{1} \and%
Jeroen Meijer\inst{2}\thanks{Supported by STW SUMBAT grant: 13859} \and%
Jaco van de Pol\inst{2} \and%
Helen Treharne\inst{3} \and%
Jorden Whitefield\inst{3}\thanks{Partly supported by EPSRC grant: EP/M506655/1}}

\authorrunning{Bendisposto et al.}

\institute{
Institut f\"{u}r Informatik, Heinrich Heine University 
D\"{u}sseldorf, Germany \\
\email{\{bendisposto@cs., p.koerner@, leuschel@cs.\}uni-duesseldorf.de} 
\and Formal Methods and Tools, University of Twente,
The Netherlands \\
\email{\{j.j.g.meijer, j.c.vandepol\}@utwente.nl} \and%
Department of Computer Science, University of Surrey, United Kingdom \\
\email{\{h.treharne, j.whitefield\}@surrey.ac.uk}}

\maketitle
%
\begin{abstract}
We present a symbolic reachability analysis approach for B that can provide a significant speedup over
 traditional explicit state model checking.
The symbolic analysis is implemented by linking \prob\ to \ltsmin, a high-performance language independent model checker.
The link is achieved via \ltsmin's \pins interface, allowing \prob to benefit from \ltsmin's analysis algorithms, while only writing a few hundred lines of glue-code, along with a bridge between \prob\ and C using  \O{}MQ.
\prob supports model checking of several formal specification languages such as B, Event-B, Z and \tla.
Our experiments are based on a wide variety of B-Method and Event-B models to demonstrate the efficiency of 
the new link.
Among the tested categories are state space generation and deadlock detection; but
 action detection and invariant checking are also feasible in principle.
In many cases we observe speedups of several orders of magnitude.
We also compare the results with other approaches for improving model checking, such as partial
 order reduction or symmetry reduction.
We thus provide a new scalable, symbolic analysis algorithm for the B-Method and Event-B, along with 
 a platform to integrate other model checking improvements via \ltsmin\ in the future.

\end{abstract}

\section{Introduction}
\label{sec:intro}

In this paper we describe the process, technique and design decisions 
we made for integrating the two tooling sets: \ltsmin and \prob.
Bicarregui et al. suggested, in a review of projects which applied formal methods~\cite{DBLP:conf/fm/BicarreguiFLW09},
that providing useable tools remained a challenge.
Recent use of the \prob tool in a rail system case study~\cite{DBLP:conf/sefm/JamesMNRSTTW13},
where model checking large industrial sized complex specifications was performed,
illustrated that there continues to be limitations with the tooling.
Model checking CSP$\parallel$B~\cite{DBLP:journals/fac/SchneiderT05} specifications in \prob was
the original motivator for this research, and based on a promising initial exploration~\cite{jorden},
this paper defines a systematic integration of the two tooling sets.


\label{sec:intro:ltsmin}

\begin{figure}[t]
    \centering
    \resizebox{\textwidth}{!}{%
        \begin{tikzpicture}[node distance=2cm]
	\large
    \begin{scope}[yshift=4cm]
        \node[node distance=2cm, text width=2.5cm] (speclan) {Specification Languages};
        \node[below of=speclan, node distance=2cm, text width=2.5cm] (p2p) {\pinspins Wrappers};
        \node[below of=p2p, node distance=2cm, text width=2.5cm] (reach) {Reachability Tools};
    \end{scope}    
    
    \begin{scope}[yshift=4cm, xshift=6.5cm]
        \matrix[nodes={draw}, column sep=1cm]{
            \node[minimum size=6mm] (mcrl2) {\mcrl}; &
            \node[minimum size=6mm] (promela) {\promela}; &
            \node[minimum size=6mm] (other) {\ldots}; &
            \node[minimum size=6mm, thick, draw=blue] (prob) {\prob}; \\
        };
        \node[right of=prob, node distance=2.5cm] (placeholder) {};
        \node[below of=placeholder, node distance=0.95cm, inner sep=0cm] {front-end};
        \node[below of=placeholder, node distance=2.9cm, inner sep=0cm] {back-end};
        
        \node[below of=mcrl2, node distance=1cm, inner sep=0cm] (a) {};
        \node[below of=promela, node distance=1cm, inner sep=0cm] (b) {};
        \node[below of=other, node distance=1cm, inner sep=0cm] (c) {};
        \node[below of=prob, node distance=1cm, inner sep=0cm] (d) {};
        
        \draw (mcrl2) edge[thick, -stealth] (a);
        \draw (promela) edge[thick, -stealth] (b);
        \draw (other) edge[thick, -stealth] (c);
        \draw (prob) edge[thick, -stealth, blue] (d);
    \end{scope}
    
    \draw[thick, dashed](.7,3cm)--(12,3cm);

    \begin{scope}[yshift=2cm, xshift=6.5cm]
        \matrix[nodes={draw}, column sep=.5cm]{
            \node[text width=1.9cm, minimum size=10mm] (cach) {Transition Caching}; &
            \node[text width=3.8cm, minimum size=10mm, thick, draw=blue] (reord) {Variable Reordering, Transition Grouping}; &
            \node[text width=2.5cm, minimum size=10mm] (por) {Partial Order Reduction}; \\
        };
        \node[above of=cach, node distance=1cm, inner sep=0cm] (a) {};
        \node[above of=reord, node distance=1cm, inner sep=0cm] (b) {};
        \node[above of=por, node distance=1cm, inner sep=0cm] (c) {};
        
        \node[below of=cach, node distance=1cm, inner sep=0cm] (d) {};
        \node[below of=reord, node distance=1cm, inner sep=0cm] (e) {};
        \node[below of=por, node distance=1cm, inner sep=0cm] (f) {};
        
        \draw (a) edge[thick, -stealth] (cach);
        \draw (b) edge[thick, -stealth, blue] (reord);
        \draw (c) edge[thick, -stealth] (por);
        
        \draw (cach) edge[thick, -stealth] (d);
        \draw (reord) edge[thick, -stealth, blue] (e);
        \draw (por) edge[thick, -stealth] (f);
    \end{scope}
    
    \draw[thick, dashed](.7,1cm)--(12,1cm);
    
    \begin{scope}[xshift=6.5cm]
        \matrix[nodes={draw}, column sep=1cm]{
            \node[minimum size=6mm] (dist) {Distributed}; &
            \node[minimum size=6mm] (mult) {Multi-core}; &
            \node[minimum size=6mm, thick, draw=blue] (sym) {Symbolic}; \\
        };
        
        \node[above of=dist, node distance=1cm, inner sep=0cm] (a) {};
        \node[above of=mult, node distance=1cm, inner sep=0cm] (b) {};
        \node[above of=sym, node distance=1cm, inner sep=0cm] (c) {};
        
        \draw (a) edge[thick, -stealth] (dist);
        \draw (b) edge[thick, -stealth] (mult);
        \draw (c) edge[thick, -stealth, blue] (sym);
    \end{scope}    
    
\end{tikzpicture}
    }
    \caption{Modular \pins architecture of \ltsmin~\cite{DBLP:conf/tacas/KantLMPBD15}}
    \label{fig:pins}
\end{figure}

\ltsmin is a high-performance language-independent model checker that 
allows numerous modelling language front-ends to be connected to various 
analysis algorithms, through a common interface, as shown in \Cref{fig:pins}. It offers a wide 
spectrum of parallel and symbolic algorithms to deal with the state 
space explosion of different verification problems. This connecting 
interface is called the \pinsfull, the basis of which consists of a 
state-vector definition, an initial state, a partitioned successor function 
(\nextstate), and labelling functions~\cite{DBLP:conf/tacas/KantLMPBD15}.
It is through \pins that we have been able to leverage the \prob tool, therefore 
allowing us to take advantage of \ltsmin's algorithmic back-ends.
In this paper we focus on the new \prob language front-end, the grouping of transitions, and
the symbolic back-end. In \Cref{sec:experiments} we also briefly discuss state variable orders.
\label{sec:intro:prob}

\prob \cite{DBLP:journals/sttt/LeuschelB08} is an animator and model checker for many different formal 
languages \cite{LeuschelPlagge:STTT10}, including the classical B-Method~\cite{DBLP:books/daglib/0015096}, Event-B~\cite{Abrial09}, CSP, CSP$\parallel$B, Z and \tla. 
\prob~can perform automatic or step by step animation of B machines, 
and can be used to systematically verify the behaviour of machines. 
The verification can identify states which do not 
meet the invariants, do not satisfy assertions or that deadlock.
At the heart of \prob\ is a constraint solver, which enables the tool
 to animate and model check high-level specifications.
The built-in model checker is a straightforward, explicit state model checker (albeit augmented with
 various features such as symmetry reduction \cite{LeMa10_311} or partial order reduction \cite{sefm14por}).
The explicit state model checker \tlc\ can also be used as a backend \cite{HansenLeuschel_ABZ14}. 

The purpose of this paper is to make use of the advanced features of the \ltsmin model checker,
 such as symbolic reachability analysis, by linking the \prob\ state exploration engine with \ltsmin.
This is achieved through a C programming interface \cite{jensBThesis} within 
the \prob tool, allowing the representation of a state to be compatible for 
\ltsmin's consumption.
In this paper the integration focuses on what is required in order to perform symbolic reachability analysis of B-Method and Event-B specifications.
The contribution of this research is a new tool integration, which can be used as a platform for further extensions.


The paper is structured as follows: \Cref{sec:pre} presents an overview of the B-Method, a running example and an illustration of definitions of transition systems used by \ltsmin.
\Cref{sec:symbolic-alg} details the symbolic reachability analysis
 and \Cref{sec:tech} outlines the implementation details. 
Section~\ref{sec:experiments} provides empirical results from performing reachability analysis benchmarking examples in \prob alone and using the new integration of the two tools.
The paper concludes in \Cref{sec:conclusion} with reflections and future work. 

\section{Preliminaries: B-Method and Transition Systems}
\label{sec:pre}
In this section we provide an overview of the B-Method and the foundations used within \ltsmin. 

\label{sec:pre:B}

\begin{figure}[tb]
\centering
\begin{tikzpicture}[->,>=stealth',bend angle=45,auto,thick]

    \node[draw=green, node distance=6cm] (1) {\tt cs=FALSE,wait=1,finished=0};
    \node[draw=green, below of=1, node distance=1.5cm] (2) {\tt cs=\underline{TRUE},wait=\underline{0},finished=0};
    \node[draw=green, right of=2, node distance=6cm] (3) {\tt cs=\underline{FALSE},wait=0,finished=0};
    \node[draw=green, right of=1, node distance=6cm] (4) {\tt cs=FALSE,wait=0,finished=\underline{1}};
    
    
    \node[inner sep=-3pt,draw=green,regular polygon,regular polygon sides=3, above of=4, node distance=.75cm, text depth=0pt] (root) {{\tt root}};
    
    \path
    (root) edge[dotted] node[left=1cm, very near start]{{\tt \bf INITIALISATION}} (1)
    (1) edge[draw=blue] node{{\tt Enter}} (2)
    (2) edge[draw=blue,below] node{{\tt Leave}} (3)
    (2) edge[draw=blue,below] node{{\tt Exit}} (4)
    (4) edge[draw=blue] node{{\tt Restart}} (1)    
    (1) edge[loop above,draw=blue] node{{\tt Leave}} (1)    
    (2) edge[loop below,draw=blue] node{{\tt CS\_Active}} (2)    
    (3) edge[loop below,draw=blue] node{{\tt Leave}} (3)    
    (4) edge[loop below,draw=blue] node{{\tt Leave}} (4);
\end{tikzpicture}
\caption{{\em MutexSimple} statespace for {\tt MAXINT=1}}
\label{fig:Mutex1States}
\end{figure}
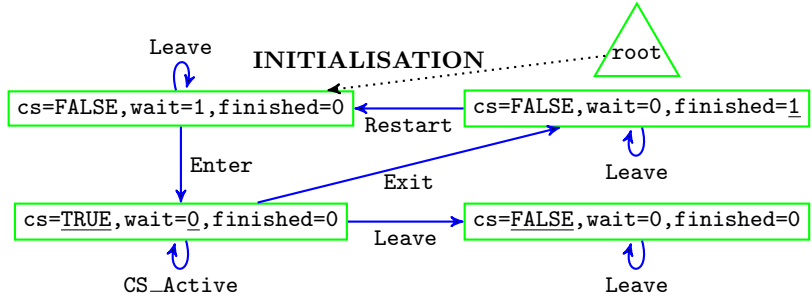

A B machine consists of a collection of clauses and a collection of operations.  The {\sc machine} clause declares the abstract machine and gives it its name. The {\sc variables} clause declares the variables that are used to carry the state information within the machine.
The {\sc invariant} clause gives the type of the variables, and more generally it also contains any other constraints on the allowable machine states.
The {\sc initialisation} clause determines the
initial state(s) of the machine.
Operations in a machine are events that change the state of a machine and can have input parameters.
Operations can be of the form {\sc SELECT} $P$ {\sc THEN} $S$ {\sc END}~where $P$ is a guard and $S$ is the action part of the operation.
The predicate $P$ must include the type of any input variables and also give  conditions on when the operation can be performed. When the guard of an operation is true then the operation is enabled and can be performed. If the guard is the simple predicate $true$ then the operation form is simplified to~{\sc BEGIN} $S$ {\sc END}.
An operation can also be of the form {\sc PRE} $P$ {\sc THEN} $S$ {\sc END}~so that the predicate is a precondition and if the operation is invoked outside its precondition then this results in a divergence (we do not illustrate this in our running example).  Finally, the action part of an operation is a {\em generalised substitution}, which can consist of one or more assignment statements (in parallel) to update the state or assign to the output variables of an operation.  Conditional statements
and nondeterministic choice statements are also permitted in the body of the operation.
The example in \Cref{fig:Bexample} illustrates the {\em MutexSimple} machine with three variables and five operations.
Its initial state is deterministic and {\em wait} is set to {\sc MAXINT}. For {\sc MAXINT}=1 we get 4 states; the state space constructed by ProB can be found in \Cref{fig:Mutex1States}. From the initial state only the guards for {\tt Enter} and {\tt Leave} are true.
Following an {\tt Enter} operation the value of the {\tt cs} variable is true which means that the guard of the {\tt CS\_Active} operation is true and the system can indicate that it is in the critical section by performing the {\tt CS\_Active} operation. 
 
The example presented could also be considered as an {\bf Event-B} example since it is a simple guarded system. We do not elaborate further on the notation of Event-B in this paper but note that the results in the subsequent sections are also applicable to Event-B.


\begin{figure}[t]
\begin{mdframed}%
\vspace{-5pt}
\lstset{
 morekeywords={MACHINE,VARIABLES,INVARIANT,INITIALISATION,OPERATIONS,BEGIN,SELECT,THEN,END}
}
\begin{lstlisting}[xleftmargin=-5pt,basicstyle=\smaller,breaklines=true,numbers=left,escapeinside={@}{@}]
MACHINE MutexSimple
VARIABLES cs, wait, finished
INVARIANT
    cs: BOOL & wait: NATURAL & finished: NATURAL
INITIALISATION cs := FALSE || wait := MAXINT || finished := 0
OPERATIONS
    Enter     = SELECT cs = FALSE & wait>0 THEN @\break@cs := TRUE || wait := wait - 1 END;
    Exit      = SELECT cs = TRUE THEN @\break@cs := FALSE || finished := finished + 1 END;
    Leave     = BEGIN cs := FALSE END;
    CS_Active = SELECT cs = TRUE THEN skip END;
    Restart   = SELECT finished>0 THEN @\break@wait := wait + 1 || finished := finished - 1 END
END
\end{lstlisting}%
\vspace{-5pt}
\end{mdframed}
\caption{{\em MutexSimple} B-Method machine example} 
\label{fig:Bexample}
\end{figure}


%

\label{sec:pre:lts}

As far as symbolic reachability analysis is concerned, a formal model is seen to denote a transition system.
\ltsmin\ adopts the following definition:

\begin{definition}[Transition System]\label{def:lts}
A Transition System (TS) is a structure $\tuple{\statese,\transrel,\initse}$, where $\statese$ is a set of states, $\transrel \subseteq \statese \times \statese$
is a transition relation and $\initse \subseteq \statese$ is a set of initial states.
Furthermore, let $\transclose$ 
be the reflexive and transitive closure of $\transrel$,
then the set of reachable states is $R = \set{s \in \statese \mid \exists s' \in \initse \suchthat s' \transclose s}$.
\end{definition}

A B-Method and Event-B model induces such a transition system: initial states are defined by the initialisation clause and
the individual operations together define the transition relation $\transrel$.
\autoref{fig:Mutex1States} shows the transition system\footnote{One subtle issue is that \ltsmin actually only supports a single initial state; this is solved by introducing the artificial \emph{root} state linked to the initial states proper. We ignore this technical issue in the paper.} for the machine in \autoref{fig:Bexample}%
.
As can be seen in \autoref{fig:Mutex1States}, the transition relation is annotated with operation names.
For symbolic reachability analysis it is actually very important that we divide the transition relation into groups,
 leading to the concept of a partitioned transition system:

\begin{definition}[Partitioned Transition System]
\label{def:pts}
A Partitioned Transition System (PTS) is a structure 
$\pts = \tuple{\states, G, \transrels, \inits}$, where
\begin{itemize}[topsep=2pt,parsep=0pt,partopsep=2pt]
\item $\states = S_1 \times \ldots \times S_\numcols$ is the set of states, which are vectors of \numcols values,
\item $G = (\to_1,\ldots,\to_{\numrows})$ is a vector of $M$ transition groups
 ${\to_i} \subseteq \states \times \states$ ($\forall 1 \leq i \leq \numrows$)
\item $\transrels = \bigcup_{i=1}^\numrows {\to_i}$ 
is the overall transition relation induced by $G$,
i.e., the union of the \numrows transition groups , and
\item $\inits \subseteq \states$ 
is the set of initial states.
\end{itemize}
We write $\vec{s} \to_i \vec{t}$ when $\tuple{\vec{s}, \vec{t}} \in {\to_i}$ 
for $1 \leq i \leq \numrows$, and
$\vec{s} \transrels \vec{t}$ when
$\tuple{\vec{s}, \vec{t}} \in \transrels$.
\end{definition}


For example $\inits = \{\tuple{FALSE,{\sc MAXINT},0}\}$ in the running example.
Note that $G$ in \Cref{def:pts} does not necessarily form a partition of $\transrels$, overlap is allowed between the individual groups.


\section{Symbolic Reachability Analysis for B}
\label{sec:symbolic-alg}

Computing the set of reachable states ($R$) of a transition system can be done efficiently
with symbolic algorithms if many transition groups $\transrel_i$ touch only a few variables.
This concept is known as event locality \cite{DBLP:journals/sttt/CiardoMS06}.
Many models of transition systems in the B-Method employ event locality.
In the B-Method event locality occurs in operations, where only a few variables are read from, or written to.
For example in \autoref{fig:Bexample} operation {\tt CS\_Active} only reads from {\tt cs} and {\tt Leave} only writes to {\tt cs}.
This event locality benefits the symbolic reachability analysis,
so that the algorithm is capable of coping with the well known state space explosion problem.
Since the B-Method employs event locality we 
build on the foundations of earlier work on \ltsmin \cite{DBLP:conf/ictac/BlomP08,DBLP:conf/hvc/MeijerKBP14}
and extend it to \prob.
To perform symbolic reachability analysis of the B-Method, \prob should provide \ltsmin
with read matrices and write matrices.
These matrices inform \ltsmin about the locality of events in the B-Method.

\ignore{***********
\begin{definition}[Read independence (OLD)]\label{def:read-independence-old}
Transition group $i$ is \emph{read-independent} from state variable $j$, if 
for all $\vec{s}, \vec{t} \in \states$ with $\vec{s} \to_i \vec{t}$, we have:
\[
  \forall r_j \exists r'_j \in S^\numcols_j \colon
  \tuple{s_1,\dotsc, r_j, \dotsc, s_\numcols} \to_i \tuple{t_1, \dotsc, r'_j, \dotsc, t_\numcols}
  \wedge r'_j\in\{r_j,t_j\},
\]
i.e., whatever value $r_j$ we plug in, the transition is still possible,
the values $t_k$ ($k \neq j$) do not depend on the value of $r_j$,
and the value of state variable $j$ is either copied ($r'_j=r_j$) or
overwritten ($r'_j=t_j$).
\end{definition}
**********}

Read independence is an important concept, it allows one to reuse the successor states computed in one state $\vec{s}$
 for all states $\vec{s'}$ which differ just by read-independent variables from $\vec{s}$, and vice versa.
\ignore{**********
\begin{definition}[Read independence (copy,overwrite)]\label{def:read-independence}
Two state vectors $\vec{s},\vec{s'}$ are equivalent on set of indexes $J$, denoted by  $\vec{s} \approx_{J} \vec{s'}$,
  iff $\forall j\in J : \vec{s}_{j} = \vec{s'}_{j}$.
We define the successors of a state $\vec{s}$ in transition group $i$, denoted by $succ_{i}(\vec{s})$, to be
  $\{t \mid \vec{s} \to_{i} \vec{t} \}$.
 
Transition group $i$ is \emph{read-overwrite independent} from state variable $j$, iff 
 $\forall \vec{s}, \vec{s'} \in S^\numcols$ such that $\vec{s} \approx_{\numcols-\{j\}} \vec{s'}$ 
 we have that
 $succ_{i}(\vec{s}) = succ_{i}(\vec{s'})$.
 
Transition group $i$ is \emph{read-copy independent} from state variable $j$, iff 
 $\forall \vec{s}, \vec{s'} \in S^\numcols$ such that $\vec{s} \approx_{\numcols-\{j\}} \vec{s'}$ 
 we have that
 $succ_{i}(\vec{s})$ = $\{ (t_{1},\ldots,t_{j-1},s_{j},t_{j+1},\ldots,t_{\numcols}) \mid \vec{t} \in succ_{i}(\vec{s'})\}$.

\end{definition}
*********}

\begin{definition}[Read independence]\label{def:read-independence}
Two state vectors $\vec{s},\vec{s'}$ are equivalent except on index $j$, denoted by  $\vec{s} \approx_{j} \vec{s'}$,
  iff $\forall k \neq j : \vec{s}_{k} = \vec{s'}_{k}$.

Transition group $i$ is \emph{read-overwrite independent} from state variable $j$, iff
$\forall \vec{s}, \vec{s'}, \vec{t} \in S^\numcols$ such that $\vec{s} \approx_j \vec{s'}$
and $\vec{s} \to_i \vec{t}$, we have that $\vec{s'}\to_i\vec{t}$.

Transition group $i$ is \emph{read-copy independent} from state variable $j$, iff
$\forall \vec{s}, \vec{s'}, \vec{t} \in S^\numcols$ such that $\vec{s}
\approx_j \vec{s'}$ and $\vec{s} \to_i \vec{t}$,
we have that $\vec{s'} \to_i (t_1,\ldots,t_{j-1},s'_{j},t_{j+1},\ldots,t_{\numcols})$.

A transition group is \emph{read independent} iff it is either read-overwrite or read-copy independent.
\end{definition}


\ignore{*** I prefer the def above
\begin{definition}[Read independence (copy,overwrite)]\label{def:read-independence}
Transition group $i$ is \emph{read-overwrite independent} from state variable $j$, iff
$\forall \vec{s}, \vec{t} \in S^\numcols$ and $s'_j\in S^j$,
if $(s_1,\ldots,s_j,\ldots,s_\numcols) \to_i (t_1,\ldots,t_j,\ldots,t_\numcols)$,
then $(s_1,\ldots,s'_j,\ldots,s_\numcols) \to_i (t_1,\ldots,t_j,\ldots,t_\numcols)$,

Transition group $i$ is \emph{read-copy independent} from state variable $j$, iff
$\forall \vec{s}, \vec{t} \in S^\numcols$ and $s'_j\in S^j$,
if $(s_1,\ldots,s_j,\ldots,s_\numcols) \to_i (t_1,\ldots,t_j,\ldots,t_\numcols)$,
then $(s_1,\ldots,s'_j,\ldots,s_\numcols) \to_i (t_1,\ldots,s'_j,\ldots,t_\numcols)$,
\end{definition}
****}

If an event never reads but may write to a variable $j$ it generally does not satisfy the above definition.
For example, the operation {\tt MayReset} = {\sc IF} $cs$ = \Btrue ~{\sc THEN} $wait$ := 0 {\sc END} \ would
 neither be read-copy nor read-overwrite independent (for state vectors with $cs=\Bfalse$ it satisfies the
  definition of the former and for $cs=\Btrue$ the latter, but neither for all state vectors).
\ltsmin\ can also deal with more liberal independence notions, but we have not yet implemented this in the present paper.
\ignore{***
 We could merge the two definitions by requiring for every equivalence class ${ \vec\{s'} \mid \vec{s} \approx_{\numcols-\{j\}} \vec{s'}\}$
  we either have read-overwrite or read-copy independence; but for a transition relation as a whole there could be a mixture.
However, this would require the model checker to know for every transition whether a variable was written or not (in particular when the variable $j$
was not modified; this could be due to a copy or an overwrite; impossible to know which). This is future work.)
***}


\begin{definition}[Write independence]\label{def:write-independence}
Transition group $i$ is \emph{write-independent} from state variable $j$, if~%
$  \forall \vec{s}, \vec{t} \in \states \colon
    \tuple{s_1, \dotsc, s_j, \dotsc, s_\numcols} \to_i \tuple{t_1, \dotsc, t_j, \dotsc, t_\numcols}
    \impl (s_j = t_j), $
i.e. state variable $j$ is never modified by transition group $i$.
\end{definition}

We illustrate the above definitions below.

\begin{definition}[Dependency Matrices]\label{def:matrix}
For a PTS $\pts = \tuple{\states,G,\transrels,\inits}$, the \emph{write matrix} is an $\numrows \times \numcols$ matrix
$WM(\pts) = WM^\pts_{\numrows \times \numcols} \in \set{0,1}^{\numrows \times \numcols}$, 
such that $(WM_{i,j} = 0) \impl$ transition group $i$ is \emph{write independent} from state variable j.
Furthermore, the \emph{read matrix} is an $\numrows \times \numcols$ matrix
$RM(\pts) = RM^\pts_{\numrows \times \numcols} \in \set{0,1}^{\numrows \times \numcols}$, 
such that $(RM_{i,j} = 0) \impl$ transition group $i$ is \emph{read independent} from state variable j.
\end{definition}

In this paper we will use sufficient syntactic conditions to ensure \Cref{def:read-independence,def:write-independence} and obtain the read and write matrix from \Cref{def:matrix}.
Indeed, we compute for every operation syntactically which variables are read from and which variables are written to.
\begin{zitemize}
\item If an operation does not write to a variable, its transition group is write independent according to \Cref{def:write-independence} and the corresponding entry in $WM$ is 0.
\item If an operation does not read a variable, its transition group is read independent according to \Cref{def:read-independence},
 unless it maybe written to  (e.g., because the assignment is in the branch of an if-then-else).
 In this case, we will mark the variable as both write and read independent.
 Also, note that when the assignment within an operation is of the form \verb+f(X) := E+ then the operation should have a read dependency on the function {\tt f} (in addition to the write dependency).
\end{zitemize}

\noindent For our example in \Cref{fig:Bexample} the syntactic read-write information is as follows:

\ignore{***
\begin{tabular}{l|l|l|l|l}
Operation &	Reads Guard &	Reads Action	& Must Write	& May Write\\
\hline
{\tt Enter}	& \{cs,wait\}	& \{cs,wait\}	& \{cs,wait\}&	$\emptyset$\\
{\tt Exit}	&\{cs\}	&\{cs,finished\}&\{cs,finished\} &$\emptyset$	\\
{\tt Leave} &	$\emptyset$	& $\emptyset$&	\{cs\}	& $\emptyset$\\
{\tt CS\_Active} &	\{cs\}	& $\emptyset$ & $\emptyset$ &$\emptyset$	\\
{\tt Restart} &	\{finished\}	&\{finished,wait\}	&\{finished,wait\}	&$\emptyset$\\
\end{tabular}
****}
\ignore{***
\begin{center}
\begin{tabular}{l|l|l|l}
Operation &	Reads	& Must Write	& May Write\\
\hline
{\tt Enter}	& \{cs,wait\}	& \{cs,wait\}&	$\emptyset$\\
{\tt Exit}	&\{cs,finished\}&\{cs,finished\} &$\emptyset$	\\
{\tt Leave} & $\emptyset$&	\{cs\}	& $\emptyset$\\
{\tt CS\_Active} &	\{cs\} & $\emptyset$ &$\emptyset$	\\
{\tt Restart} &\{finished,wait\}	&\{finished,wait\}	&$\emptyset$\\
\end{tabular}
\end{center}
****}

\begin{figure}
\vspace{-25pt}
\centering
\subfloat[Read matrix ($RM$)]{{%
$\bbordermatrix{
                    & \mathtt{cs}   & \mathtt{wait} & \mathtt{finished} \cr
\mathtt{Enter}      & \mathbf{1}    & \mathbf{1}    & 0                 \cr
\mathtt{Exit}       & \mathbf{1}    & 0             & \mathbf{1}        \cr
\mathtt{Leave}      & 0             & 0             & 0                 \cr
\mathtt{CS\_Active}   & \mathbf{1}    & 0             & 0                 \cr
\mathtt{Restart}   & 0    & \mathbf{1}             & \mathbf{1}                 \cr
}$}}%
\quad%
\subfloat[Write matrix ($WM$)]{{%
$\bbordermatrix{
                    & \mathtt{cs}   & \mathtt{wait} & \mathtt{finished} \cr
\mathtt{Enter}      & \mathbf{1}    & \mathbf{1}    & 0                 \cr
\mathtt{Exit}       & \mathbf{1}    & 0             & \mathbf{1}        \cr
\mathtt{Leave}      & \mathbf{1}    & 0             & 0                 \cr
\mathtt{CS\_Active}   & 0             & 0             & 0                 \cr
\mathtt{Restart}   & 0    & \mathbf{1}             & \mathbf{1}                 \cr
}$}}%
\vspace{-5pt}
\caption{Dependency matrices}%
\label{fig:deps}%
\end{figure}
\vspace{-20pt}

From the matrices we can infer if a variable is read-copy or read-overwrite independent: a variable that is read independent and not written to (i.e., write independent) is read-copy independent, otherwise it is read-overwrite independent.
 We can thus infer that:
\begin{zitemize}
\item the transition group of {\tt Enter} is read-copy and write independent on {\tt finished}.
\item {\tt Exit} is read-copy and write independent on {\tt wait}.
\item {\tt Leave} is read-copy and write independent on {\tt wait} and {\tt finished} and read-overwrite independent on {\tt cs}.
\item {\tt CS\_Active} is read-copy and write independent on {\tt wait} and {\tt finished} and write independent on {\tt cs} (but not read-independent on {\tt cs}).
\item {\tt Leave} is read-copy and write independent on {\tt cs}.
\end{zitemize}
\subsection{Exploration Algorithm}\label{ssec:explore}
%
We now present the core of the symbolic reachbility analysis algorithm of \ltsmin.
\Cref{alg:smc} computes the set of reachable states $R$ (represented as a decision diagram) and it uses the
independence information to minimise the number of next state computations that have to be carried out,
i.e., re-using the next states $\{t\mid\vec{s} \transrel_{i} \vec{t}\}$ computed for a single state $\vec{s}$
for many other states $\vec{s'}$ according to \Cref{def:read-independence,def:write-independence}.
\Cref{alg:smc} will, while it keeps finding new states, expand the partial transition relation with potential successor states,
and apply the expanded relation to the set of new states.
 
\begin{figure}[tb]
\centering
\includegraphics[width=.7\textwidth]{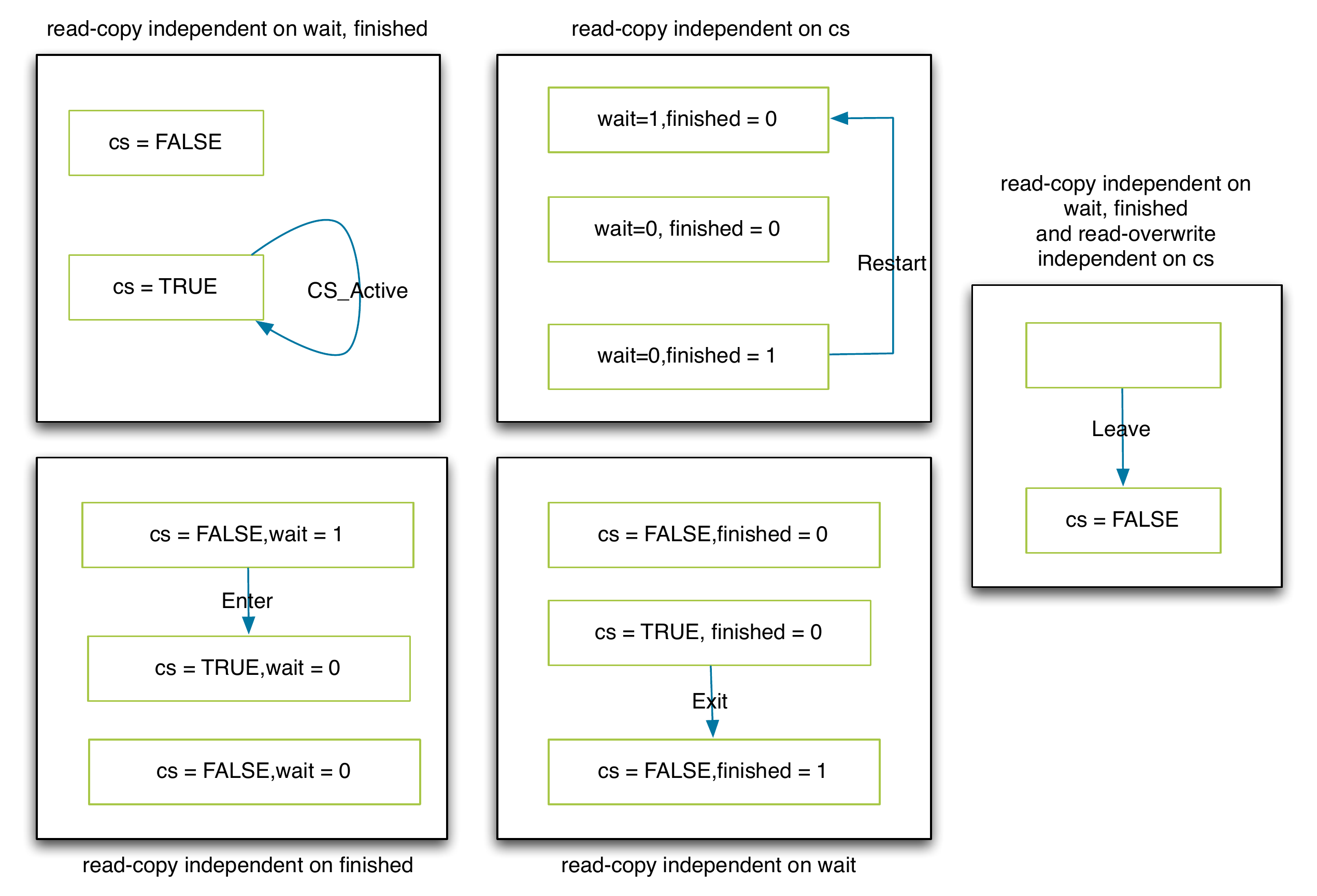}
\caption{MutexSimple, operations computed on their projected state space}
\label{fig:Mutex1Ops}
\end{figure}

Four key functions that make \Cref{alg:smc} highly performant are the following.%
\footnote{We refrain from giving their formal definitions; they can be found in \cite{DBLP:conf/hvc/MeijerKBP14}.}
The (1) \emph{read projection} $\pi^r_i = \pi^{RM}_i$ and (2) \emph{write projection} $\pi^w_i = \pi^{WM}_i$
take as argument a state vector and produce a state vector restricted
to the read and write dependent variables of group $i$, respectively.
Furthermore these function are extended to apply to sets directly,
e.g., given the examples in \Cref{fig:Bexample,fig:deps}, a read projection for \texttt{Leave} is
$\pi^r_3(\set{\tuple{\underline{FALSE},0,0},\tuple{\underline{FALSE},0,1},\tuple{\underline{FALSE},1,0}}) = \set{\tuple{\underline{FALSE}}}$.
This is illustrated in \Cref{fig:Mutex1Ops} and used at \Cref{alg:line-project} in \Cref{alg:smc:learn-trans}.
The read projection prevents \ltsmin from doing two unnecessary next state calls to \prob, 
since \texttt{Leave} is \emph{read-copy} independent on \texttt{wait} and \texttt{finished}.

The function (3) $\nextstate_i$ takes a read projected state and projects (with $\pi^w_i$) all successor states of transition group $i$.
The partial transition relation $\transrelexpr^\mathrm{p}_i$ is learned on the fly,
and $\nextstate_i$ is used to expand $\transrelexpr^\mathrm{p}_i$.
An example next state call for \texttt{Enter} is $\nextstate_1(\tuple{FALSE,1}) = \set{\tuple{TRUE,0}}$.

Lastly, (4) $\textsc{next}$ takes a set of states, a partial transition relation, a row of the read and write matrix and outputs a set of successor states.
For example, applying the partial relation of {\tt Enter} to the initial state yields
$\textsc{next}(\set{\tuple{\underline{FALSE},\underline{1},0}},\set{\tuple{\tuple{FALSE,1},\tuple{TRUE,0}}},\tuple{1,1,0},\tuple{1,1,0}) = \set{\tuple{\underline{TRUE},\underline{0},0}}$.
Note that in this example {\tt Enter} is \emph{read-copy} independent on {\tt finished}
and thus \textsc{next} will copy its value from the initial state.

The usage of these four key functions is also illustrated in \Cref{fig:MutexAlgStep}.
The figure shows that first the projection is done for \texttt{Enter},
then $\transrelexpr^\mathrm{p}_i$ is expanded with a $\nextstate_i$ call,
lastly relation $\transrelexpr^\mathrm{p}_i$ is applied to the initial state,
producing the first successor state.

\begin{figure}
\vspace{-25pt}
\begin{tabular}{p{.55\linewidth}@{\hspace{.5em}}p{.43\linewidth}}
\begin{minipage}[t]{\linewidth}
\begin{algorithm}[H]
\caption{\textsc{ReachBreadthFirst}
}
\label{alg:smc}
\scriptsize
\SetKwInOut{Input}{Input}
\SetKwInOut{Output}{Output}
\Input{$\inits \subseteq S^\numcols, \mathrm{M} \in \natnumset, RM, WM$}
\Output{The set of reachable states $\mathcal{R}$}
$\mathcal{R} \gets \inits$; $\mathcal{L} \gets \mathcal{R}$\; \label{alg:initstart}
\lFor{$1 \leq i \leq \mathrm{M}$}{%
    $\mathcal{R}_i^\mathrm{p} \gets \emptyset$;
    $\transrelexpr_i^\mathrm{p} \gets \emptyset$%
} \label{alg:initend}
\While{$\mathcal{L} \neq \emptyset$} { \label{line:smc:outer-while}
    \textsc{LearnTrans}(); $\mathcal{N} \gets \emptyset$\;    
    \For{$1 \leq i \leq \mathrm{M}$} {
        $\mathcal{N} \gets \mathcal{N} \cup \textsc{next}(\mathcal{L}, \transrelexpr_i^\p, RM_i, WM_i)$\;\label{alg:callnext}%
    }
    $\mathcal{L} \gets \mathcal{N} \setminus \mathcal{R}$; $\mathcal{R} \gets \mathcal{R} \cup \mathcal{N}$\; 
} \label{alg:whileend}
\Return{$\mathcal{R}$}
\end{algorithm}
\end{minipage}
&
\begin{minipage}[t]{\linewidth}
\begin{algorithm}[H]
\caption{\textsc{LearnTrans}}
\label{alg:smc:learn-trans}
\SetKwInOut{Output}{Description}
\Output{Extends $\transrelexpr_i^\p$}
    \For{$1 \leq i \leq \mathrm{M}$} {
        $\mathcal{L}^\mathrm{p} \gets \pi^r_i(\mathcal{L})$\; \label{alg:line-project}
        \For{$s^\mathrm{p} \in \mathcal{L}^\mathrm{p} \setminus \mathcal{R}_i^\mathrm{p} $} {
            $\transrelexpr_i^\mathrm{p} \gets \transrelexpr_i^\mathrm{p} \cup \{\tuple{s^\mathrm{p},d^\mathrm{p}} \mid$ \label{alg:line-rel-update1}\\
            $d^\mathrm{p} \in  \nextstate_i(s^\mathrm{p})\}$\;\label{alg:line-rel-update2}
        }
        $\mathcal{R}_i^\mathrm{p} \gets \mathcal{R}_i^\mathrm{p} \cup \mathcal{L}^\mathrm{p}$\;
    }
\end{algorithm}
\end{minipage}
\end{tabular}
\vspace{-20pt}
\end{figure}

 \begin{figure}[tb]
\centering
\includegraphics[width=.5\textwidth]{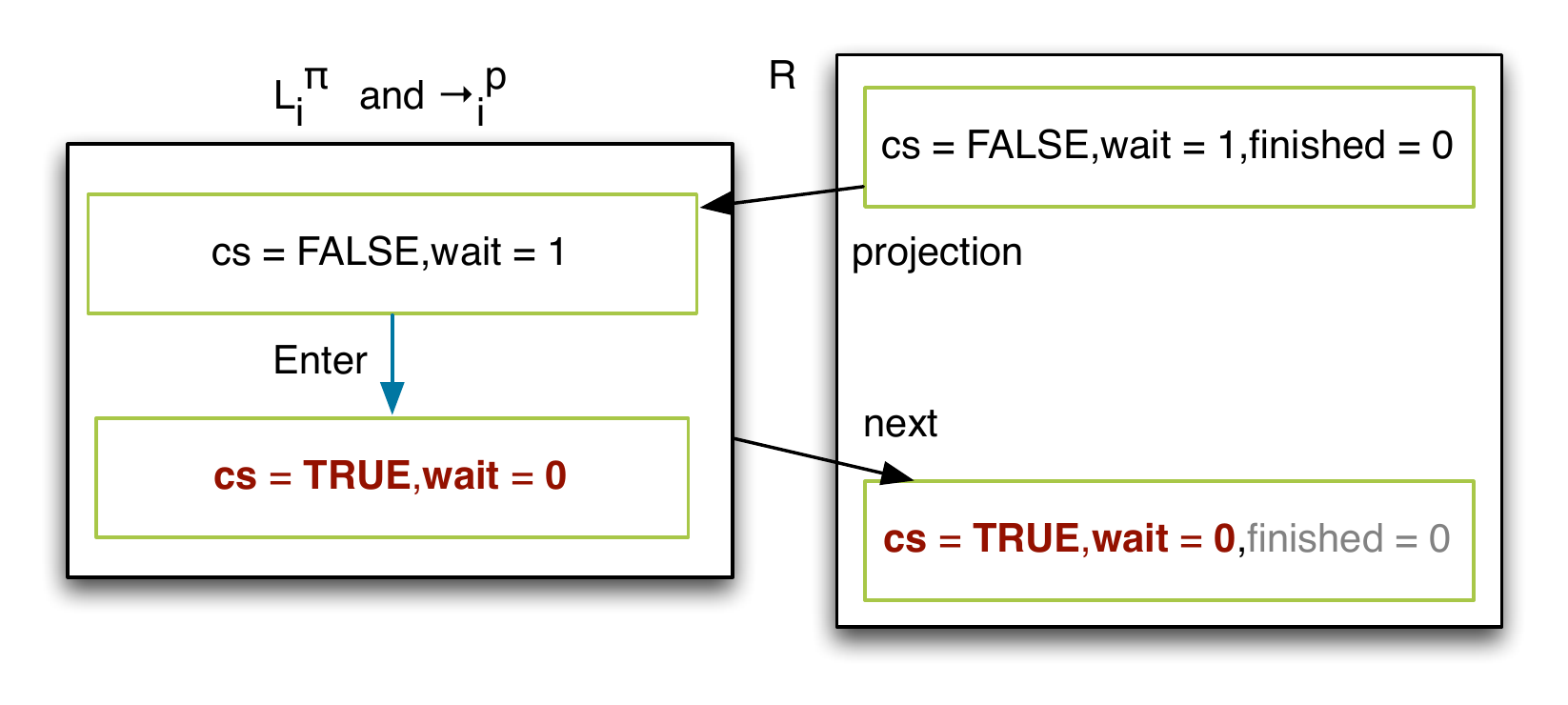}
\caption{Illustrating one iteration of \Cref{alg:smc} for MutexSimple and \texttt{Enter} operation}
\label{fig:MutexAlgStep}
\end{figure}

\Cref{fig:Mutex1Ops} shows for each operation the transition relation $\transrelexpr^\mathrm{p}_{i}$ and the projected states
on which they are computed.
\Cref{def:read-independence} ensures that the projected state space shown in \Cref{fig:Mutex1Ops}
can be used to compute the effect of each of these operations for the \emph{entire} state space (using \texttt{next}).

\subsection{List Decision Diagrams}

\begin{figure}
\vspace{-30pt}
\centering
\resizebox{.7\textwidth}{!}{
\subfloat[Variables]{%
{\includegraphics[height=125pt]{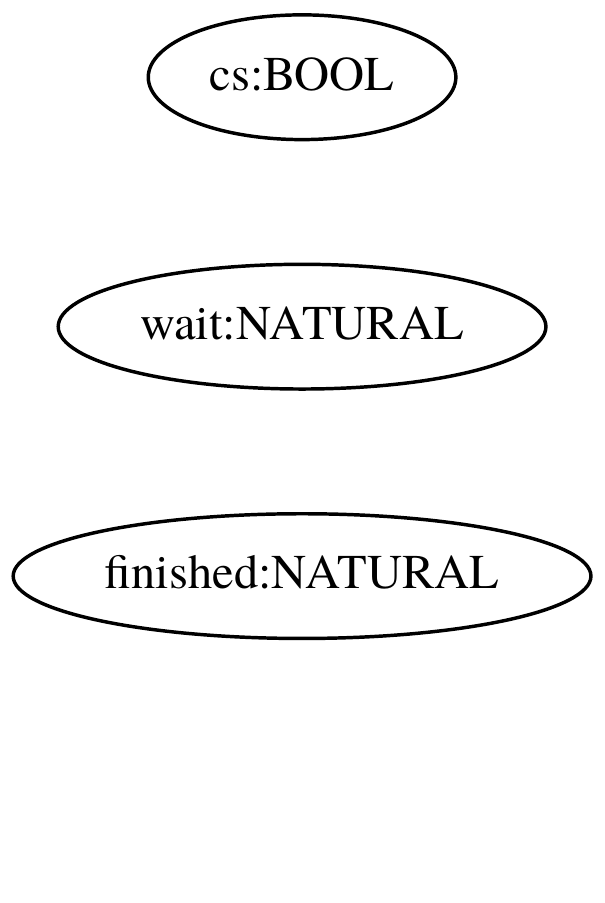}\label{fig:ldd:names} }}%
\quad%
\subfloat[Iteration 1]{%
\hspace*{20pt}
{\includegraphics[height=125pt]{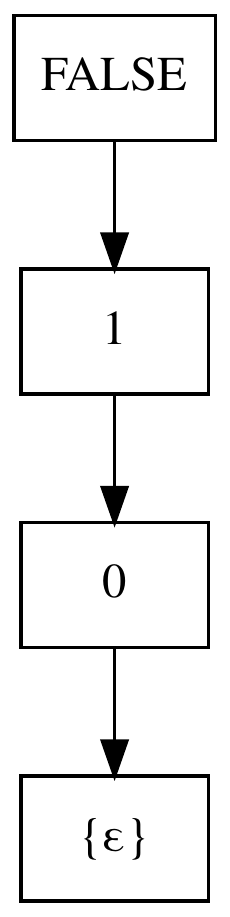}\label{fig:ldd:l1} }}%
\quad%
\subfloat[Iteration 2]{%
{\includegraphics[height=125pt]{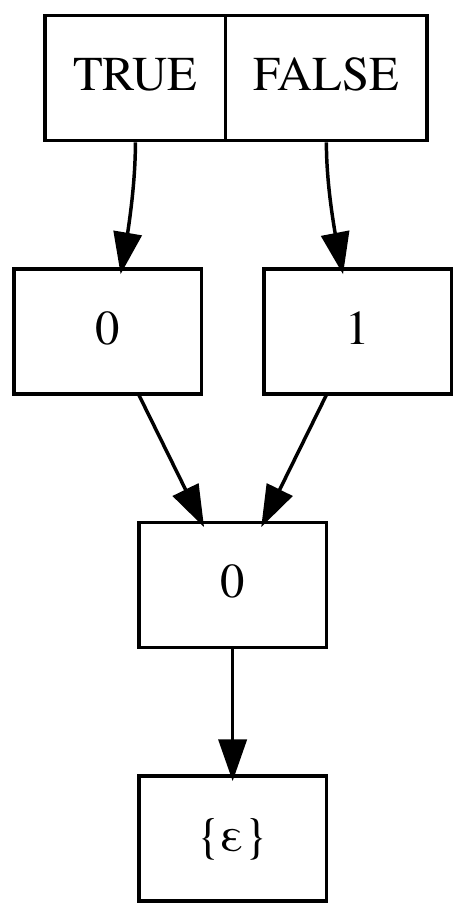}\label{fig:ldd:l2} }}%
\quad%
\subfloat[Iteration 3]{%
{\includegraphics[height=125pt]{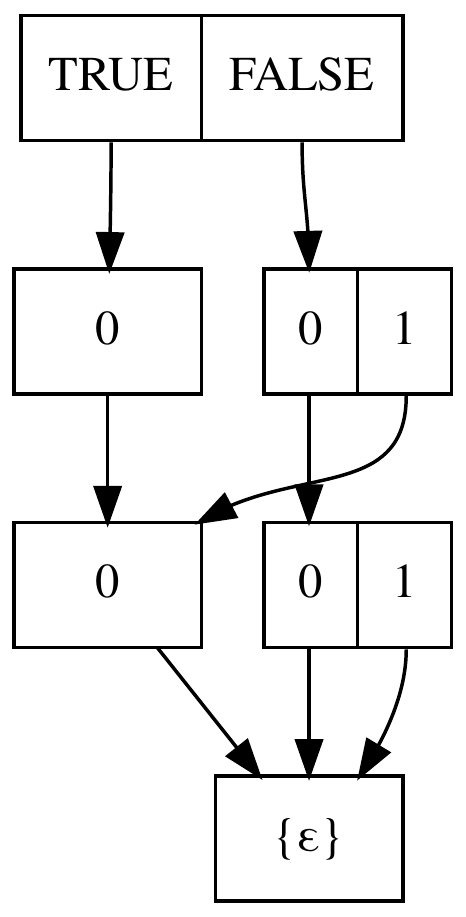}\label{fig:ldd:l3} }}}%
\caption{LDDs of the reachable states}%
\label{fig:Mutex1LDD}%
\vspace{-5pt}
\end{figure}
\vspace{-20pt}

The symbolic reachability algorithm in \Cref{ssec:explore} uses List Decision Diagrams (LDDs) to store the reachable states and transition relations.
Similar to a Binary Decision Diagram, an LDD \cite{DBLP:conf/ictac/BlomP08} represents a set of vectors.
Due to the sharing of state vectors within an LDD, the memory usage can be very low, even for very large state spaces.
Three example LDDs for the running example are given in \Cref{fig:Mutex1LDD}.
The LDDs represent the set of reachable states $\mathcal{R}$ in 
\Cref{alg:smc} at each iteration of \Cref{line:smc:outer-while}.
In an LDD every path from the top left node to $\set{\epsilon}$ is a state, e.g., the initial state $\tuple{FALSE,1,0}$ in \Cref{fig:ldd:l1}.
A node in an LDD represents a unique set of (sub) vectors,
e.g., $\set{\epsilon}$ represents the set of zero-length vectors and
the right-most $0$ of variable {\tt wait} in \Cref{fig:ldd:l3} encodes the set $\set{\tuple{0,0},\tuple{0,1},\tuple{1,0}}$.
\Cref{fig:ldd:l2} shows that firing {\tt Enter} will add $\tuple{TRUE,0,0}$ to $R$.
In \Cref{fig:ldd:l3} $\tuple{FALSE,0,0}$ and $\tuple{FALSE,0,1}$ are added to $R$, by firing {\tt Leave} and {\tt Exit} respectively.
The benefit of using LDDs for state storage is due to the sharing of state vectors.
For example, the subvector $\tuple{FALSE}$ of the states
\set{$\tuple{FALSE,0,0}, \tuple{FALSE,0,1}, \tuple{FALSE,1,0}, \tuple{FALSE,1,1}$}
in iteration 3 is encoded in the LDD with a single node.
For bigger state spaces the sharing can be huge; resulting in a low memory footprint
for the reachability algorithm.

\subsection{Performance: \nextstate function}
\label{sec:background:nuf}
There are two big differences of \Cref{alg:smc}
with classical explicit state model checking as used by \prob \cite{DBLP:journals/sttt/LeuschelB08}.
First, the state space is represented using an LDD datastructure, which enables sharing amongst states.
Second, independence is used to apply the \nextstate function not state by state, but for
 entire sets of states in one go.
For each of the 4 states in \Cref{fig:Mutex1States}, the explicit model checking algorithm of
 \prob\ would check whether each of the 5 operations is enabled; resulting in 20 next-state calls.
With \ltsmin's symbolic reachability \Cref{alg:smc}, only 12 \nextstate calls are made.
This is shown in the following table, where + means enabled, - means disabled, and C means that \ltsmin has reused the results of a previous call to \prob.

\begin{figure}
\vspace{-20pt}
\centering
\begin{tabular}{c||c|c|c||c|c|c|c|c}
State\# & cs & wait & finished & {\tt Enter} & {\tt Exit} & {\tt Leave} & {\tt CS\_Active} & {\tt Restart}\\ 
\hline
1 & FALSE & 1 & 0 & + & C & C & C & -\\ 
2 & TRUE & 0 & 0 & - & + & + & + & -\\ 
3 & FALSE & 0 & 0 & - & - & C & - & C\\ 
4 & FALSE & 0 & 1 & C & - & C & C & +\\ 
\end{tabular}
\vspace{-20pt}
\end{figure}

If we initialise {\tt wait} with $\mathrm{MAXINT}=500$, the state space has 251,002 states.
The runtime with \prob is 70 seconds, with \ltsmin{}+\prob 48 seconds and \ltsmin performs only 6012 \nextstate calls.
The example does not have a lot of concurrency and uses only simple data structures (and thus the overhead
of the \ltsmin's \prob front-end is more of a factor compared to the runtime of ProB for computing successor states);
other examples will show greater speedups (see \Cref{sec:experiments}).
But the purpose of this example is to illustrate the principles.

\section{Technical Aspects and Implementation}
\label{sec:tech} 

\begin{wrapfigure}[7]{r}{0.7\textwidth}
\vspace{-8pt}
\centering
\includegraphics[width=.7\textwidth]{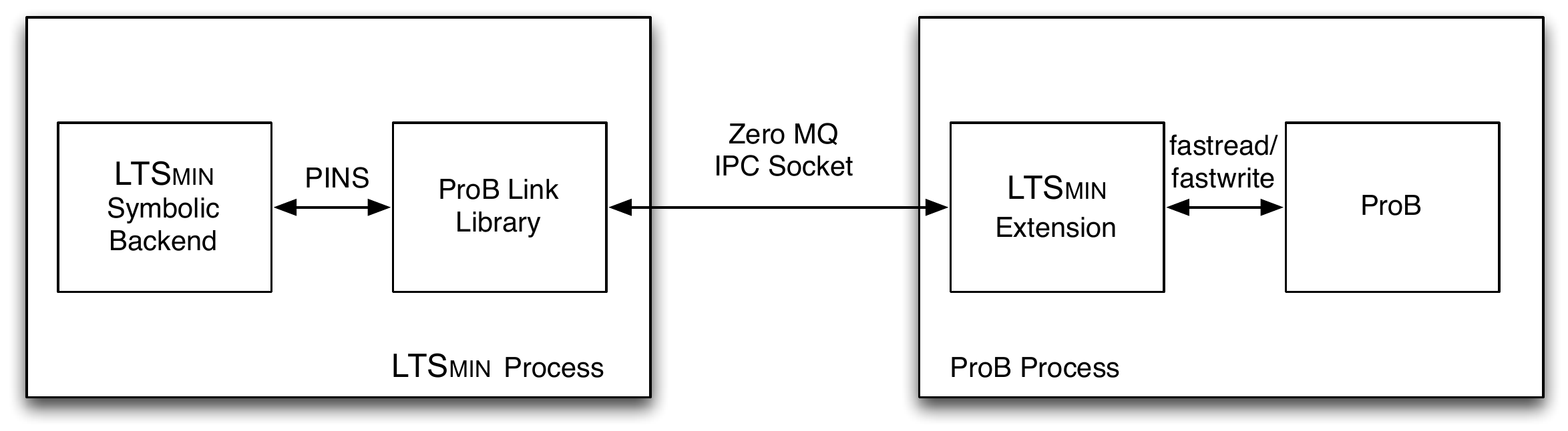}
\caption{High level design showing the integration} 
\label{fig:integration}
\vspace{-5pt}
\end{wrapfigure}

We used a distributed approach to integrate \prob and \ltsmin. 
Both tools are stand-alone applications, so a direct integration, i.e., turning one of the tools into a shared library would require considerable effort. 
We therefore added extensions to both tools that convert the data formats and use sockets to communicate with each other. 
A high level view of the integration is shown in \Cref{fig:integration}.  
We use the  \O{}MQ 
  \cite{0mq} library for communication. 
\O{}MQ is oriented around message queues and can be used as both, a networking library with very high throughput and as a concurrency framework. 
We have chosen \O{}MQ because it worked very well in previous work~\cite{jensBThesis}. 
Although we do not (yet) have to care about concurrency in this work, the reactor abstraction provided by \O{}MQ was very handy in the \prob extension. It allows to implement a server that receives and processes messages without much effort.
The communication is always initiated by \ltsmin; it sends a message and blocks until it receives the answer from \prob. 
 
We usually run both tools on a single computer using interprocess (IPC) sockets, but it is only a matter of configuration to run the tools on different machines using TCP sockets.
We currently only support Linux and Mac OS.
The communication protocol is straightforward. Reachability analysis is initiated from \ltsmin by sending an initialisation packet.
 \prob answers with a message containing the relevant static information about a model, such as the dependency matrices that \ltsmin requires (see \Cref{sec:symbolic-alg}). 
 
Each matrix is encoded as a 2-dimensional array, which is not optimal for a sparse matrix but is not an issue because we only send the matrices once.
The packet also contains the list of variables, their types, the list of transition groups, and the initial state.
 
States are represented as a list of so called {\bf chunks}. 
A chunk is one of the elements in the state vector according to \Cref{def:pts}. 
In the case of B, each chunk is a value of one of the state variables. 
Because \ltsmin will not look inside the chunks, we simply use the binary representation of \prob's Prolog term that represents the value of a variable. 
This has the advantage, that \prob does not have to keep information about the state space. 
It can always recover a state from the chunks that are sent by \ltsmin. 
The transition groups correspond to B operations as explained in \Cref{sec:pre:lts}. 
Like chunks the transition groups are only used as names in \ltsmin. 




Once the initial setup is done, \ltsmin will start to ask \prob for successor states for {\bf specific} transition groups. 
It will send a next-state message containing a state and a transition group. 
The state, that \ltsmin sends is a list of chunks and \prob's \ltsmin extension can directly consume them and construct a Prolog term that internally represents a state.
Using this constructed state and the transition group, the extension will then ask \prob for all successor states. 
The result is a list of Prolog terms, each representing a successor state. 
The extension transforms the list of states into a list of lists of chunks and sends them back to \ltsmin. 
This is repeated until \ltsmin has explored all necessary states and sends a termination signal.

The next-state messaging is similar to \Fref{fig:MutexAlgStep}, the projection is achieved by replacing all read independent variables by a dummy value. 
\section{Experiments}
\label{sec:experiments}

To demonstrate that the combination of \prob\ and \ltsmin\ improves the performance of the reachability analysis and deadlock detection compared with the standalone version of \prob, we use a wide range of B and Event-B models. 
Our benchmark suite contains puzzles (e.g., towers of Hanoi) as well as specifications of protocols (e.g., Needham-Schroeder), algorithms (e.g., Simpson's four slot algorithm) and industrial specifications (e.g., a choreography model by SAP, a cruise control system by Volvo and a fault tolerant automatic train protection system by Siemens).%
\footnote{More detailed descriptions can be found in \condtext{\cite{LtsMinTR}}{Appendix~\ref{appendix-models}}.}



The experiments were run on Ubuntu 15.10 64-bit, with 8 GB RAM, 120 GB SSD and an Intel Sandybridge Mobile i5 2520M 2.50 GHz Dual core. 
The version of \prob used in this paper is 1.5.1-beta3, and \ltsmin tag LTSminProBiFM2016
\footnote{The software and models can be found online at \url{https://github.com/utwente-fmt/ProB-LTSmin-iFM16}.}

Figure~\ref{fig:ProBbenchmarksJorden} summarises a selection of the experiments that we ran. The last two models are Event-B models.
In these experiments we used Breadth-First Search (BFS) and looked for deadlocks.
A deadlock was found only for the Philosophers model (this is also why there are no next state call statistics for this model).
The table also contains the number of next state calls for \prob\ reachability analysis on its own
 and when called from \ltsmin's symbolic reachability analysis algorithm (i.e., our new integration see \Cref{sec:background:nuf}) without deadlock checking.
One can clearly see that we obtain a considerable reduction in wallclock time.
The \prob\ time is the walltime of the \prob reachability analysis and initial state computation and does not include parsing and loading.
The \ltsmin CPU time column shows how much time is spent in the \ltsmin side of the symbolic reachability analysis  algorithm.
The \ltsmin wall time shows the total walltime, and this also contains the time spent in the communication layer and waiting for the \prob\ process to compute the next states.
To compare the benefit of our new algorithm we compute the speedup of the walltime in the last column by dividing the \prob walltime from column 5 with the \ltsmin walltime in column 7.

We can see that for some of the smaller models the overhead of setting up \ltsmin does not pay off.
However, for all larger models, except for the Train1\_Lukas\_POR model considerable speedups were obtained.

\begin{figure}[htb]
\vspace{-15pt}
\begin{scriptsize}
\begin{tabular}{l|r|r|r|r|r|r|r|r}
Benchmark & Events &  States & \prob & \ltsmin & \prob & \ltsmin & \ltsmin & Speedup\\
          &  &              &  Nxt St &  NxtSt & Wall &  CPU & Wall  & \\
          &  &              &  Calls &  Calls &  (ms) &  (ms) & (ms) & \\
\hline
CAN\_BUS & 21 & 132600  & 2784560 & 3534 & 122850 & 660 & 1590 & 77.264\\
ConcurrentCounters & 4 & 110813  & 443249 & 113032 & 21820 & 2760 & 13820 & 1.579\\
Cruise\_finite1 & 26 & 1361 & 35361 & 1667 & 2900 & 100 & 1020 & 2.843\\
file\_system & 8 & 698 & 5577 & 1198 & 1900 & 180 & 4660 & 0.41\\ 
MutexSimple & 5 & 10 & 46 & 26 & 10 & 10 & 190 & 0.053\\
Philosophers & 5 &  &  &  & 480 & 40 & 590 & 0.814 \\
SiemensMiniPilot\_Abrial0 & 9 & 181  & 1621 & 182 & 100 & 20 & 260 & 0.385\\
Simpson\_Four\_Slot & 9 & 46658  & 419906 & 2089 & 17310 & 200 & 860 & 20.128\\
Train1\_Lukas\_POR & 8 & 24637 & 197082 & 101441 & 33660 & 6480 & 50260 & 0.670\\
nota & 11 & 80719  & 887899 & 588 & 287970 & 130 & 660 & 436.318\\
pkeyprot2 & 10 & 4412 & 44111 & 2004 & 22190 & 210 & 1710 & 12.977\\
\hline
Ref5\_Switch\_mch & 38 & 29861  & 1134681 & 1281 & 160600 & 490 & 1260 & 127.460\\
obsw\_M001 & 21 & 589279 & 12374779 & 23406 & 2051320 & 1620 & 12420 & 165.163\\
\end{tabular}
\end{scriptsize}
\caption{B and Event-B Machines, with BFS and deadlock detection}
\label{fig:ProBbenchmarksJorden}
\vspace{-5pt}
\end{figure}
\vspace{-20pt}

A major result we achieved with non default settings for \ltsmin, is for elevator12.eventb.
This model is not listed in \Cref{fig:ProBbenchmarksJorden}, because \prob runs out of memory on the hardware configuration used for this experiment. 
\ltsmin computed in 34 seconds, with 96,523 \nextstate calls, that the model has 1,852,655,841 states.
As reachability algorithm we chose chaining \cite{DBLP:conf/apn/RoigCP95},
and to compute a better variable order, we ran Sloan's bandwidth reduction algorithm \cite{NME:NME1620281111} on the dependency matrix.

As far as memory consumption is concerned; when performing reachability analysis on CAN\_BUS, the \prob process consumes 370 MB real memory, while
the \ltsmin process consumes 633 MB, with the default settings.
With the default settings \ltsmin will allocate $2^{22}$ elements ($\approx$ 100 MB) for the node table and $2^{24}$ elements ($\approx$ 500 MB) for the operations cache.
If we choose a smaller node table and operations cache for the LDD package (both $2^{18}$ elements), \ltsmin consumes only 22 MB.
The default settings for \ltsmin are geared towards larger symbolic state spaces than that of CAN\_BUS.
The default node table and cache are too big for CAN\_BUS, and thus not completely filled during reachability.

We have also run our new symbolic reachability analysis on Z and \tla\ models.
For example, we successfully validated the video rental Z model from \cite{DBLP:conf/asm/DerrickNS08}.
For 2 persons and 2 titles and maximum stock level of 4, \ltsmin\ generates the 23009 states in 1.75 seconds
 compared to 52.4 seconds with \prob alone. The model contained useless constants; after removing them \prob\ runs in 1.6 seconds; the runtime of \ltsmin stays unchanged.
We were unable to use the output of z2sal \cite{DBLP:conf/asm/DerrickNS08} using SAL \cite{SAL-SRI} and its symbolic model checker for comparison. 

In summary, \Cref{fig:ProBbenchmarksJorden} shows that for several non-trivial B and Event-B models, considerable improvements can
be obtained using the symbolic reachability analysis technique described in this paper. 

{\bf Alternate Approaches}
\label{sec:experiments-alternate}
Other techniques for improving model checking for B-Method and Event-B models have been developed
 and evaluated in the recent years.
We have run a further set of experiments using a selection of those methods; the complete results can be found
 in \condtext{\cite{LtsMinTR}}{Appendix~\ref{appendix:more-experiments}}.
For technical reasons, the experiments were run on different hardware than above, a MacBook Air with 2.2 GHz i7 processor and 8GB of RAM.
We summarise the findings here and compare the results with our new symbolic model checking algorithm.

\begin{center}
\begin{scriptsize}
\begin{tabular}{|l|rr|rr|rr|r|}
\hline
Benchmark  & \multicolumn{2}{c|}{\prob\ POR}  & \multicolumn{2}{c|}{\prob\ Hash }  & \multicolumn{2}{c|}{\tlc} &  \prob\ no opt.\\
 & ms & Speedup & ms & Speedup & sec & Speedup & ms\\
 \hline
CAN\_BUS  & 138720 & 0.80 & 98390 & 1.12 & 3 & 37 & 110400\\
ConcurrentCounters  & 50 & 345.8 & 18400 & 1.06 & 1 & 17 & 17290\\
file\_system  & 2380 & 0.37 & 210 & 4.24 & 29 & 0.03 & 890\\
Simpson\_Four\_Slot  & 20860 & 0.70 & 9550 & 1.52 & 1 & 15 & 14530\\
Train1\_Lukas\_POR  & 34030 & 0.75 & 28930 & 0.88 & 4 & 6 & 25740\\
nota  & 490 & 509.22 & 14780 & 16.88 & 10 & 25 & 249520\\
Ref5\_Switch\_mch  & 215160 & 0.59 & 124500 & 1.01 & 6 & 21 & 126170\\ 
obsw\_M001  & 2150520 & 0.80 & 76190 & 22.53 & 55 & 31 & 1716770\\  
\hline
\end{tabular}
\end{scriptsize}
\end{center}

 The authors in~\cite{HansenLeuschel_ABZ14} presented a translation from the B-Method to \tla, with the goal of using the
{\bf \tlc\ } model checker \cite{DBLP:conf/charme/YuML99} as backend.
\tlc\ has no constraints solving capabilities, and as such that it can only deal with lower level models.
On the other hand, its execution can be considerably faster than \prob, and its explicit state model checking engine
 (which stores fingerprints) is very efficient.
On the downside, there is a small probability that fingerprint collisions can occur.
The experiments show that \tlc\ does not deal well with benchmark programs which require constraint
 solving (graph isomorphism, JobsPuzzle, \ldots), running up to three orders of magnitude slower
 than \prob\ or \ltsmin\ with \prob.
However, it does deal very well with lower level  models, e.g., it is faster than \ltsmin\ for ConcurrentCounters. 
For many benchmark models, even those not requiring constraint solving, our symbolic reachability analysis
 is faster.
For example, for the nota example, \tlc\ runs in about 10 seconds---faster than \prob\ without any optimisation---but
 slower than \ltsmin\ by less than a second.

{\bf Symmetry} reduction \cite{LeMa10_311} can be very useful; but exponential improvements usually occur only on academic examples.
Here we have experimented with the hash marker symmetry reduction, 
 which is \prob's fastest symmetry method, but is generally not guaranteed to explore all states.
The method gives the best results for certain models (e.g., file\_system). 
But for several of the larger, industrial examples shown above, its benefit is not of the same scale as \ltsmin.
In future, we will investigate combining \prob's symmetry reduction with the new \ltsmin algorithm.

We have also experimented with {\bf partial order reduction}.
\cite{sefm14por} uses a semantic preprocessing phase to determine independence
  (different from our purely syntactic determination; see \Cref{sec:symbolic-alg}).
As such, it can induce a slow down for some examples where this does not pay off  
 (e.g., file\_system). 
\prob's partial order reduction obtains the best times for certain models with a large degree of concurrency (ConcurrentCounters, SiemensMiniPilot\_Abrial, and nota). 
However, once we start doing invariant checking, \cite{sefm14por}
 does not scale nearly as well (e.g., it takes 134 seconds instead of 0.5 seconds for \ltsmin checking the nota model).
But even without invariant checking, there are plenty examples where the symbolic reachability analysis approach is
 better (e.g., Cruise\_finite1, Philosophers, Simpson\_Four\_Slot and almost two orders of magnitude for CAN\_BUS).
In summary:
\begin{zitemize}
\item \tlc\ is good for models not requiring constraint solving. It is a very efficient, explicit state model checker.
However, models often have to be rewritten (such as CAN\_BUS), and there is a small chance of having fingerprint collisions.
\item Symmetry reduction excels when models make use of deferred sets.
 However, the hash marker method is not guaranteed to explore all states.
\item Partial order reduction is very good for models with a large degree of concurrency.
However, it can cause slow downs and is less suited for invariant checking.
\item The new symbolic reachability analysis algorithm deals well with concurrency and is by far the fastest method for certain larger, industrial models, such as CAN\_BUS, obsw\_M001, elevator12, the ABZ landing gear model or Abrial's mechanical press. 
\ltsmin is currently the only tool set that uses a symbolic representation of the state space that is connected to \prob.
\end{zitemize}

\section{More Related Work, Future Work and Conclusion}
\label{sec:conclusion}


We have already evaluated the use of \tlc\ \cite{DBLP:conf/charme/YuML99} for model checking B.
Another explicit state model checker for B has been presented in \cite{DBLP:conf/icfem/MatosFM09}, which uses lazy enumeration.
Symbolic model checking \cite{Bur92a} has been used for railway applications in \cite{DBLP:conf/isola/Winter12}.
The best known symbolic model checker is probably SMV \cite{McMillan1993}, which uses a low-level input language.
Some comparisons between using SMV and \prob\ have been conducted in \cite{DBLP:conf/saicsit/HorneP08}, where models were translated by hand.
For abstract state machines there is the AsmetaSMV \cite{DBLP:conf/asm/ArcainiGR10} tool, which automatically translates ASM specifications to SMV.
It is our impression that the translation often leads to a considerable blowup of the model, encoded in SMV's low-level language, also affecting performance.
We did one experiment on a Tic-Tac-Toe model provided for AsmetaSMV: NuSMV 2.6 took over 13 seconds to find a configuration where the cross player wins; \prob\ (without \ltsmin) took 0.2 seconds model checking time for the same property on a similar B model.
Another experiment involved puzzle 3 of the RushHour game: \prob\ solves this in 5 seconds, while NuSMV still had not found a solution after 120 minutes.

Other symbolic model checkers that perform comparable well to \ltsmin include \textsc{Marcie} \cite{HRS13} and \textsc{PetriDotNet}\cite{DBLP:conf/tacas/MolnarDVB15}.

The paper provides a stable architectural link between \prob and \ltsmin that can be extended.
First, we plan to provide \ltsmin\ with more fine-grained information about the models, both statically and dynamically.
Dynamically, \prob\ will transmit to \ltsmin which variables have actually been written by an operation, enabling a more extensive independence notion to be used.
Statically, \prob\ will transmit the individual guards of operations and provide variable read
 matrices for the guards.
We will also transmit the individual invariants in the same manner, to enable analysis of the invariants.
(It is actually already possible to check invariants using the present integration, simply by encoding invariants
 as operations. We have done so with success for some of the examples, e.g., the nota from Section~\ref{sec:experiments}.)
When \prob transmits individual guards, we also hope to use the guard-based partial order optimisations of \ltsmin \cite{laarman2014:guard} and enable LTL model checking with \ltsmin.



These future directions will strengthen the capability of the 
verification tools and hence further encourage the application 
of formal methods within industry as identified in~\cite{DBLP:conf/fm/BicarreguiFLW09}, for example to support complex railway systems verification in CSP$\parallel$B. This will require both more fine-grained static and dynamic information.

In summary, we have presented a new scalable, symbolic analysis algorithm for the B-Method and Event-B, along with 
 a platform to integrate other model checking improvements via \ltsmin\ in the future.

\bibliographystyle{splncs03}
\bibliography{references}

\condtext{ 
}{

 \appendix
 
 \section{Models for experiments (for referees)} \label{appendix-models}
 
 These model descriptions will be uploaded to a website, also containing the publicly available models.
 
 The classical B models shown in the tables in Section~\ref{sec:experiments} are:
\begin{footnotesize}
\begin{itemize}
 
 \item  {\bf CAN\_BUS}\\A model of a controller area network (CAN) bus developed by Colley.
 
 \item  {\bf ConcurrentCounters}\\
  This is a very simple model with three concurrent counters that can be incremented independently.
  
 \item  {\bf Cruise\_finite1}\\
 Volvo Vehicle Function. The B specification machine has
 15 variables, 
  550 lines of B specification,
 and 26 operations.  The invariant consists of 40 conjuncts.
 This B specification was developed by Volvo
 as part of the European Commission IST Project PUSSEE (IST-2000-30103).
 
 \item {\bf file\_system} is a simple model of a file system with users, groups and access control.
 
 \item  {\bf MutexSimple}\\
  This is the model shown in Fig.~\ref{fig:Bexample} of the paper.
  
 \item  {\bf Philospohers}\\
  This is a B model of the well-known dining philosphers problem.
 
\item {\bf SiemensMiniPilot\_Abrial0}\\
This Siemens Mini Pilot was developed within the Deploy Project.
It is a specification of a fault-tolerant automatic train protection system that ensures
that only one train is allowed on a part of a track at a time.
The model contains a single refinement level and rather complex invariants.
The model was translated to classical B from Event-B.

 \item {\bf Simpson\_Four\_Slot}\\
  A model of Simpson's four slot algorithm. This B model only represents the individual steps of the algorithm.
  It is intended to be used in conjunction with a CSP model to describe the sequencing of the steps.
  Here, the B model on its own is model checked (thus leading to invariant violations).
  
 \item  {\bf Train1\_Lukas\_POR}\\
 This is the first level of refinement of the railway interlocking model in Chapter 17 of \cite{Abrial09}.
 It uses a simplified topology and routes are released immediately when all blocks are free (to reduce the state space).
 
 \item  {\bf nota}\\ 
 A model developed by Nokia within the RODIN Project\footnote{{\tt http://rodin.cs.ncl.ac.uk/}} for the validation and verification of Nokia's
NoTA hardware platform; see \cite{DBLP:conf/b/Oliver07}.
  
 \item {\bf pkeyprot2 (Needham-Schroeder)}\\
  The Needham-Schroeder public key protocol is an authentication protocol for creating a secure connection over a public network~\cite{NedSchroed78}.
  The model consists of a network with the two normal users called Alice and Bob, an attacker named Eve and the keyserver.
  The first version of this protocol, developed in 1978, contains an error which was found in 1995 by Lowe. 
  This model is a slightly simplified version (reducing the messages sent by Eve).
\end{itemize}
\end{footnotesize}

 The Event-B models shown in the tables in Section~\ref{sec:experiments} are:
\begin{footnotesize}
\begin{itemize}
\item {\bf Ref5\_Switch\_mch} 
This is the fifth level of refinement of a solution \cite{abz14casestudyjournal} to the ABZ'14 landing gear challenge.

\item {\bf obsw1\_M001}\\
The Space Systems Finland example is a model of a subsystem used for the ESA BepiColombo mission.
 The model is a specification of parts of the BepiColombo On-Board software, that contains a core software and two subsystems used for telecommand and telemetry of the scientific experiments, the Solar Intensity X-ray and particle Spectrometer (SIXS) and the Mercury Imaging X-ray Spectrometer (MIXS). 
The model was a mini pilot of the Deploy project.

 \item {\bf elevator12}\\
 This is the twelfth refinement of an elevator model by ETH Z\"urich.
 
\end{itemize}
\end{footnotesize}

Here are additional experiments that were run, but whose results are not shown in the main paper (due to page limit restrictions).
\begin{footnotesize}
\begin{itemize}

 \item {\bf BlocksWorldGeneric6}\\
  A model of blocksworld, with six blocks; the goal being to put all blocks in the right-order on top of each other.

 \item {\bf Echo}\\
 The Echo algorithm 
 is designed to find the shortest paths in a network topology.

 \item {\bf Hanoi6}\\ 
 The well-known towers of Hanoi puzzle with 6 discs.
  
 \item {\bf scheduler\_bztt}\\
 The process scheduler from \cite{LegeardEtAl:FME02}. 

 \item {\bf RushHour}\\
  The Rush Hour puzzle.%
\footnote{See {\tt http://en.wikipedia.org/wiki/Rush\_Hour\_(board\_game)}.}
 This is the hardest puzzle (number 40 in the regular version of the game).
  The shortest solution needs 83 moves; here we have explored the full state space of 4782 states and 29890 transitions.
 
 \item  {\bf Cansell\_Contention}\\ 
 A Firewire-Leader election protocol by Dominique Cansell. 
 
\item {\bf CXCC0}\\
CXCC (Cooperative Crosslayer Congestion Control) ~\cite{Scheuermann:CXCC} 
 is a cross-layer approach to prevent congestion in wireless networks. 
The invariants used in the model are rather complex.

 \item  {\bf press\_7b\_mch}\\ 
 This is the seventh level of refinement of Abrial's model of a mechanical press. 
 \end{itemize}
\end{footnotesize}

\newpage
\section{More Experiments: Partial order reduction, Symmetry and TLC (for referees)}
\label{appendix:more-experiments}

We have run a further set of experiments using a selection of alternate methods for improving model checking for B and Event-B models.
These experiments were run on a MacBook Air with 2.2 GHz i7 processor.
For the final version of the paper, these experiments will be re-run on the same hardware as in Sect.~\ref{sec:experiments}, for all relevant \prob\ and \ltsmin settings. 
The experiments, models and setup will be put onto a website for open access.

\begin{zitemize}
\item \cite{HansenLeuschel_ABZ14} presented a translation from classical B to TLA+, with the goal of using the
{\bf \tlc\ }.
We used this to run \tlc\ with one worker, no invariant, no deadlock and no assertion and no LTL checking.
The table in Sect.~\ref{sec:experiments} of the paper contains just the \tlc\ model checking time (as measured by \tlc\ to a one second accuracy).
However, the table below includes the full runtime; including JVM startup and parsing.
The factor removes 1 second from runtime; in the final version of the paper we will obtain more precise model checking times for \tlc.
It can also not be directly applied to Event-B or Z models (requiring a translation to classical B first); hence
the numbers for the Event-B models are missing.
\item Hash marker symmetry reduction \cite{LeMa10_311}.
 It  is the fastest symmetry method, but is generally not guaranteed to explore all states
The time is the walltime of the model checking and initial state computation and does not include parsing and loading.
\item partial order reduction method of  \cite{sefm14por}.
It uses a semantic preprocessing phase to determine independence.
The time is the walltime of the model checking and initial state computation and does not include parsing and loading.
\end{zitemize}
An analysis of these results can be found in Section~\ref{sec:experiments-alternate}.
First, we present the models also shown in Figure~\ref{fig:ProBbenchmarksJorden} in Section~\ref{sec:experiments-alternate}:

\begin{scriptsize}
\begin{tabular}{|l|rr|rr|rr|r|}
\hline
Benchmark  & POR &  & Hash &  & TLC &  &  No opt.\\
 & ms & Factor & ms & Factor & ms & Factor & ms\\
 \hline
CAN\_BUS  & 138720 & 1.26 & 98390 & 0.89 & 5340 & 0.04 & 110400\\
ConcurrentCounters  & 50 & 0.00 & 18400 & 1.06 & 2130 & 0.07 & 17290\\
Cruise\_finite1  & 5060 & 1.92 & 2330 & 0.89 & 2950 & 0.74 & 2630\\
MutexSimple  & 30 & 1.50 & 30 & 1.50 & 1660 & 33.00 & 20\\
Philosophers  & 4180 & 52.25 & 100 & 1.25 & 1690 & 8.63 & 80\\
SiemensMiniPilot\_Abrial\_0  & 70 & 0.70 & 100 & 1.00 & 1440 & 4.40 & 100\\
Simpson\_Four\_Slot  & 20860 & 1.44 & 9550 & 0.66 & 2670 & 0.11 & 14530\\
Train1\_Lukas\_POR  & 34030 & 1.32 & 28930 & 1.12 & 6650 & 0.22 & 25740\\
nota  & 490 & 0.00 & 14780 & 0.06 & 11780 & 0.04 & 249520\\
pkeyprot2  & 24140 & 1.29 & 17990 & 0.96 & 1190 & 0.01 & 18770\\
Ref5\_Switch\_mch  & 215160 & 1.71 & 124500 & 0.99 & - &  & 126170\\
obsw\_M001  & 2150520 & 1.25 & 76190 & 0.04 & - &  & 1716770\\
\hline
\end{tabular}
\end{scriptsize}

~\\
These are some additional experiments that we performed; \tlc\ could not run on the RushHour model due to a bug in \tlc.

\begin{scriptsize}
\begin{tabular}{|l|rr|rr|rr|r|}
\hline
Benchmark  & POR &  & Hash &  & TLC &  &  No opt.\\
 & ms & Factor & ms & Factor & ms & Factor & ms\\
 \hline
BlocksWorldGeneric6  & 2980 & 0.94 & 950 & 0.30 & 2240 & 0.39 & 3160\\
CAN\_BUS\_normalized  & 105460 & 1.43 & 77090 & 1.05 & 4590 & 0.05 & 73530\\
CSM  & 700 & 1.37 & 50 & 0.10 & 1790 & 1.55 & 510\\
EchoAlg  & 1000 & 3.57 & 280 & 1.00 & 2730 & 6.18 & 280\\
GardnerSwitchingPuzzle\_v2  & 1390 & 1.78 & 780 & 1.00 & 4510 & 4.50 & 780\\
GraphIsoMedium  & 50 & 0.83 & 110 & 1.83 & 40360 & 656.00 & 60\\
Hanoi6  & 440 & 0.98 & 440 & 0.98 & 1540 & 1.20 & 450\\
JobsPuzzle  & 100 & 1.11 & 80 & 0.89 & 32410 & 349.00 & 90\\
NQueens  & 40 & 1.00 & 30 & 0.75 & 20550 & 488.75 & 40\\
PhilRing  & 1500 & 1.95 & 120 & 0.16 & 1820 & 1.06 & 770\\
Reading  & 120 & 1.00 & 40 & 0.33 & 1470 & 3.92 & 120\\
brueckenproblem  & 530 & 1.89 & 300 & 1.07 & 2150 & 4.11 & 280\\
fahrzeugverwaltung2  & 2880 & 1.18 & 90 & 0.04 & 2840 & 0.75 & 2440\\
file\_system  & 2380 & 2.67 & 210 & 0.24 & 28600 & 31.01 & 890\\
scheduler\_bztt  & 260 & 3.25 & 60 & 0.75 & 1340 & 4.25 & 80\\
RushHour\_v2\_TLC  & 18670 & 1.25 & 16520 & 1.11 & err & - & 14920\\
cont0  & 330 & 0.65 & 320 & 0.63 & 1290 & 0.57 & 510\\
Model\_4\_NoDeadlock\_v6  & 850 & 0.50 & 590 & 0.35 & - &  & 1710\\
VM\_4\_mch  & 1050 & 1.62 & 860 & 1.32 & - &  & 650\\
cxcc0  & 410 & 5.86 & 70 & 1.00 & - &  & 70\\
press\_7b\_mch  & 9960 & 1.23 & 2500 & 0.31 & - &  & 8090\\
scheduler  & 50 & 1.25 & 30 & 0.75 & - &  & 40\\
\hline
\end{tabular}
\end{scriptsize}

\ignore{******

Flooding symmetry: precise method; generates full statespace like symbolic mc.
Note: for TokenRing: the previous methods did not compute the full state space (because of MAX\_INITIALISATIONS).
Flooding does. Hence the huge increase in runtime and number of states.
Nxt state calls computation is wrong; only one element per symmetry class will be evaluated.

\begin{footnotesize}
\begin{tabular}{l|r|r|r|r}
Benchmark & States & Transitions & Events  & ProB (ms) \\
\hline
BlocksWorldGeneric6 & 4052 & 7139 & 2 &  980 \\
CAN\_BUS & 132600 & 340266 & 21 & 100160 \\
CAN\_BUS\_normalized & 132600 & 340266 & 21 & 76040 \\
CSM & 77 & 209 & 13 & 520 \\
ConcurrentCounters & 110813 & 325003 & 4 & 17650 \\
Cruise\_finite1 & 1361 & 25696 & 26 & 2360 \\
EchoAlg & 246 & 469 & 4  & 270 \\
GardnerSwitchingPuzzle\_v2 & 212 & 510 & 2  & 830 \\
GraphIsoMedium & 72 & 81 & 1  & 60 \\
Hanoi6 & 731 & 2186 & 1  & 500 \\
JobsPuzzle & 5 & 4 & 0  & 90 \\
MutexSimple & 10 & 23 & 5  & 20 \\
NQueens & 9 & 40 & 1 & 30 \\
PhilRing & 5857 & 6149 & 3  & 290 \\
Philosophers & 130 & 225 & 5  & 100 \\
Reading & 115 & 252 & 5  & 60 \\
SiemensMiniPilot\_Abrial\_mch\_0 & 181 & 991 & 9  & 100 \\
Simpson\_Four\_Slot & 46658 & 80114 & 9  & 10340 \\
TokenRing & 19355439 & 19360030 & 4  & 3074330 \\
nota & 80719 & 166213 & 11  & 17600 \\ 
pkeyprot2 & 4412 & 62900 & 10 & 18360 \\ 
\end{tabular}
\end{footnotesize}
***}

}

\end{document}